\documentclass[aps,prb,twocolumn,superscriptaddress,floatfix]{revtex4}

\usepackage{graphicx,color}
\usepackage{dcolumn}
\usepackage{bm}
\usepackage{stmaryrd}
\usepackage{latexsym}
\usepackage{amssymb}
\usepackage{amsfonts}
\usepackage{amsmath}
\usepackage{epstopdf}

\hfuzz=\maxdimen
\begin{document}


\title{Strong anisotropy effect in iron-based superconductor CaFe$_{0.882}$Co$_{0.118}$AsF}

\author{Yonghui Ma}
\affiliation{State Key Laboratory of Functional Materials for
Informatics, Shanghai Institute of Microsystem and Information
Technology, Chinese Academy of Sciences, Shanghai 200050,
China}\affiliation{CAS Center for Excellence in Superconducting
Electronics(CENSE), Shanghai 200050, China}\affiliation{University
of Chinese Academy of Science, Beijing 100049, China}

\author{Qiucheng Ji}
\affiliation{State Key Laboratory of Functional Materials for
Informatics, Shanghai Institute of Microsystem and Information
Technology, Chinese Academy of Sciences, Shanghai 200050,
China}\affiliation{CAS Center for Excellence in Superconducting
Electronics(CENSE), Shanghai 200050, China}\affiliation{University
of Chinese Academy of Science, Beijing 100049, China}

\author{Kangkang Hu}\affiliation{State Key Laboratory of Functional Materials for
Informatics, Shanghai Institute of Microsystem and Information
Technology, Chinese Academy of Sciences, Shanghai 200050, China}
\affiliation{Shanghai Key Laboratory of High Temperature
Superconductors, Shanghai University, Shanghai 200444, China}

\author{Bo Gao}
\affiliation{State Key Laboratory of Functional Materials for
Informatics, Shanghai Institute of Microsystem and Information
Technology, Chinese Academy of Sciences, Shanghai 200050, China}

\author{Wei Li}
\email[]{liwei@mail.sim.ac.cn} \affiliation{State Key Laboratory of
Functional Materials for Informatics, Shanghai Institute of
Microsystem and Information Technology, Chinese Academy of Sciences,
Shanghai 200050, China}\affiliation{CAS Center for Excellence in
Superconducting Electronics(CENSE), Shanghai 200050,
China}\affiliation{Department of Physics and State Key Laboratory of
Surface Physics, Fudan University, Shanghai 200433, China}

\author{Gang Mu}
\email[]{mugang@mail.sim.ac.cn} \affiliation{State Key Laboratory of
Functional Materials for Informatics, Shanghai Institute of
Microsystem and Information Technology, Chinese Academy of Sciences,
Shanghai 200050, China}\affiliation{CAS Center for Excellence in
Superconducting Electronics(CENSE), Shanghai 200050, China}

\author{Xiaoming Xie}
\affiliation{State Key Laboratory of Functional Materials for
Informatics, Shanghai Institute of Microsystem and Information
Technology, Chinese Academy of Sciences, Shanghai 200050,
China}\affiliation{CAS Center for Excellence in Superconducting
Electronics(CENSE), Shanghai 200050, China}

\begin{abstract}
The anisotropy of the Fe-based superconductors is much smaller than
that of the cuprates and the theoretical calculations. A credible
understanding for this experimental fact is still lacking up to now.
Here we experimentally study the magnetic-field-angle dependence of
electronic resistivity in the superconducting phase of iron-based
superconductor CaFe$_{0.882}$Co$_{0.118}$AsF, and find the strongest
anisotropy effect of the upper critical field among the iron-based
superconductors based on the framework of Ginzburg-Landau theory.
The evidences of energy band structure and charge density
distribution from electronic structure calculations demonstrate that
the observed strong anisotropic effect mainly comes from the strong
ionic bonding in between the ions of Ca$^{2+}$ and F$^-$, which
weakens the interlayer coupling between the layers of FeAs and CaF.
This finding provides a significant insight into the nature of
experimentally observed strong anisotropic effect of electronic
resistivity, and also paves an avenue to design exotic two
dimensional artificial unconventional superconductors in future.

\end{abstract}

\maketitle

\section{introduction}
Since the discovery of superconductivity of 26 K in fluorine doped
quaternary compound LaFeAsO,~\cite{LaFeAsO} the studies of the
mechanism of iron-based superconductors (FeSCs) have evoked enormous
interests across the community of superconductivity and material
science. Huge experiments have revealed that the iron-based
superconductors belong to a family of an unconventional pairing
mechanism within layered FeAs(Se). The information of upper critical
field $H_{c2}$ and its anisotropy in superconductivity can be used
as a fingerprint to understand the unconventional superconducting
mechanism and to promote the practical applications
.~\cite{gamma-1,gamma-2,8} In addition, bidimensionality, which is
represented by a rather high anisotropy between the $ab$-plane and
$c$ direction, is believed to be a very important factor for the
occurrence of strong correlation physics and even high-$T_c$
superconductivity~\cite{CDW1,CDW2,SDW1,Cuprates,Chu}. Generally
speaking, the anisotropy parameter $\gamma$ of the FeSCs was found
to be much smaller than that of the cuprates and the theoretical
calculations.~\cite{NdFeAsO0.82F0.18,10-4-8,10,11,12,13,Singh1,Cuprates}
Although the Pauli-limit effects have been taken into account in
well understanding the low temperature behaviors of
$H_{c2}$,~\cite{Pauli-limiting1,Pauli-limiting2,Pauli-limiting3} the
nature of such small value of anisotropy parameter $\gamma$ of FeScs
remains unclear in the vicinity of superconducting (SC) transition
temperature $T_c$. Interestingly, it has been found that the values
of $\ln(\gamma^2)$ is linearly proportional to the distance $d$ of
adjacent conducting layers for both the cuprates supeconductors and
FeSCs,~\cite{Cuprates,10-4-8} which is understandable since a
thicker insulating block layer weakens the interlayer hybridization,
as a result the anisotropy will be enhanced. Quantitatively, the
value $\gamma^2$ of FeSCs has a magnitude typically two orders lower
than that of cuprates superconductors with the same $d$.

Recently, high-quality single crystals of CaFeAsF and the Co-doped
compounds with the size above 1 mm were grown successfully by our
group using a new flux CaAs,~\cite{20,CaFeAsF-Co} which facilitates
our research on the anisotropy effect of this system. Although the
anisotropy parameter of the penetration depth were investigated by
torque measurements previously,~\cite{torque1,torque2} it was found
that the value of $\gamma$ obtained from London penetration depth
experiment is quite different from that based on the upper critical
field in FeSCs.~\cite{penetration-depth1,penetration-depth2} The
nature of such difference remains unclear. Motivated by this issue,
we systematically study the anisotropy of the superconducting single
crystals CaFe$_{0.882}$Co$_{0.118}$AsF based on the upper critical
fields. The magnetic-field-angular dependence of electronic
resistivity in the superconducting state is measured and the results
are found to follow the scaling law of the anisotropic
Ginzburg-Landau (G-L) theory. The fitted anisotropy parameter is
clearly larger than that of other FeSCs, and reaches the order of
magnitude of the theoretical estimations.~\cite{Singh1} Based on the
charge density distributions calculations, it is found that this
strange phenomena mainly stems from the nature of the strong ionic
bonding in between the ions of Ca$^{2+}$ and F$^-$, which weakens
the interlayer coupling between the FeAs and CaF layers.

\section{Experimental details and Calculations}
The CaFe$_{1-x}$Co$_{x}$AsF single crystals were grown using the
CaAs self-flux method.~\cite{20,CaFeAsF-Co} The crystal structure
and lattice constants of the materials were examined by a DX-2700
type powder x-ray diffractometer using Cu K$_\alpha$ radiation. The
electronic resistivity (including the magnetic field angle-resolved
electronic resistivity) were measured using a four probe technique
on the physical property measurement system (Quantum Design, PPMS)
with magnetic field up to 9 T. The angle $\varphi\ $ was varied from
0$^{\circ}$ to 180$^{\circ}$, where $\varphi$ = 0$^{\circ}$
indicates the orientation with the magnetic field parallel to the
$c$-axis of the sample. In addition, the current was applied always
perpendicular to the direction of magnetic field, as shown
schematically in the inset of Fig.~\ref{fig2}(a).

The first-principles calculations presented in this work were
performed using the all-electron full potential linear augmented
plane wave plus local orbitals (FP-LAPW + lo) method~\cite{LAPW} as
implemented in the WIEN2K code.~\cite{wien2k} The
exchange-correlation potential was calculated using the generalized
gradient approximation (GGA) as proposed by Pedrew, Burke, and
Ernzerhof.~\cite{PBE} These calculations were performed using the
experimental crystal structure~\cite{20}, as shown in Fig. 1(a).

\begin{figure}
\includegraphics[width=9cm]{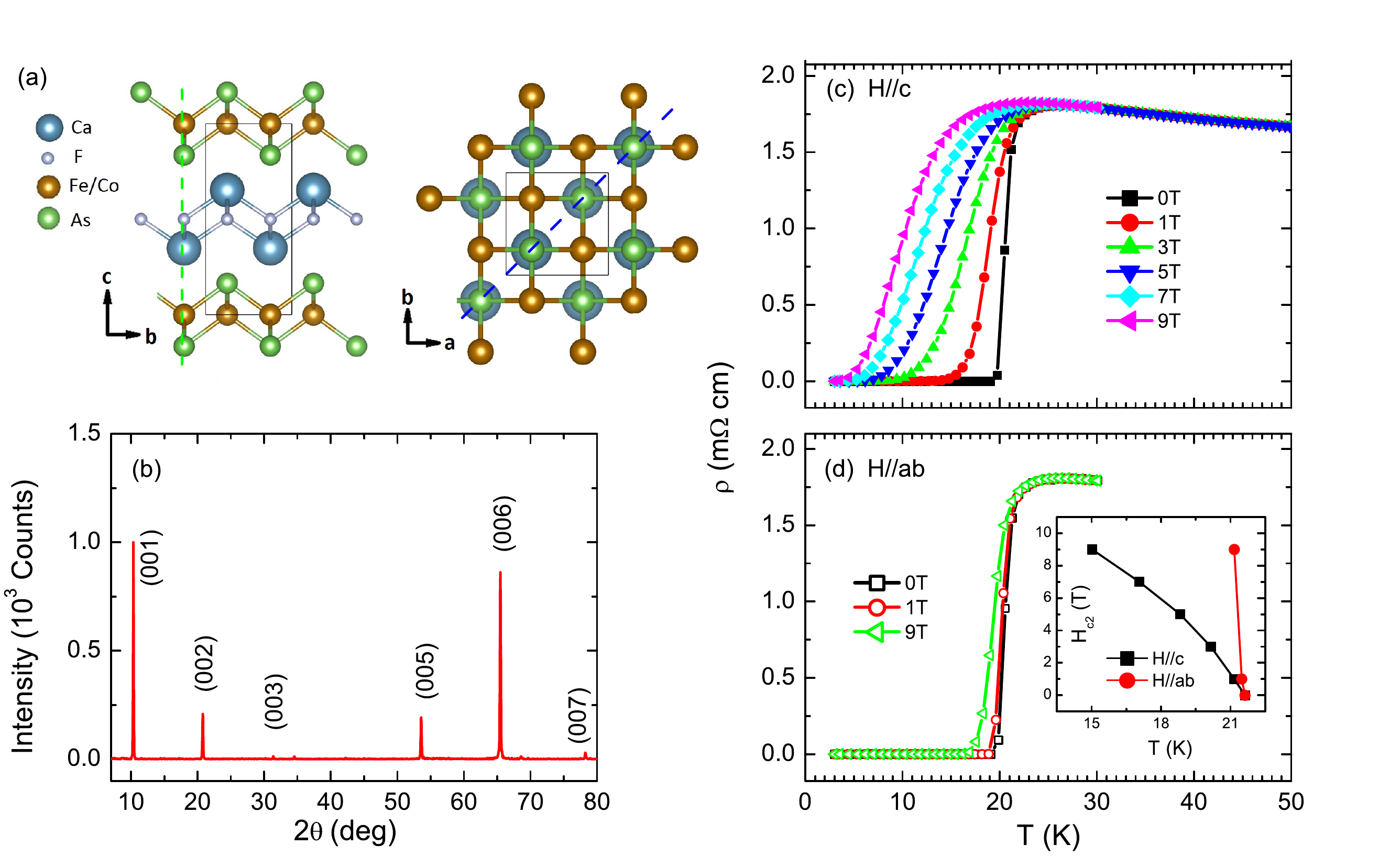}
\caption {(a) The schematic crystal structure of
CaFe$_{1-x}$Co$_{x}$AsF on (100) and (001) planes. The green and
blue dashed lines denote the (010) and (110) planes where the charge
density distribution was calculated (see Fig. 3). (b) X-ray
diffraction pattern measured on the single crystal
CaFe$_{0.882}$Co$_{0.118}$AsF with the incident of x-ray on the
$ab$-plane. (c)-(d) The electronic resistivity as a function of
temperature under the magnetic field up to 9 T with $H//c$ and
$H//ab$, respectively. The inset of (d) shows the upper critical
fields H$_{c2}$ as a function of temperature for two different
orientations.} \label{fig1}
\end{figure}

\section{Results}
The x-ray was incident on the $ab$-plane of the sample when carrying
out the x-ray diffraction measurements. The diffraction pattern is
shown in Fig.~\ref{fig1}(b), which follows the tetragonal
ZrCuSiAs-type structure. It is important to point out that only the
sharp peaks along the (00$l$) orientation are observed, suggesting a
high $c$-axis orientation. The actual doping level of cobalt was
determined to be 0.118 by the energy dispersive x-ray spectroscopy.
Detailed analysis and discussions about the crystal structure and
the chemical composition have been reported in our previous
paper.~\cite{CaFeAsF-Co} Temperature dependence of electronic
resistivity is shown in Fig.~\ref{fig1}(c) and (d). Under zero
field, the sample exhibits a sharp SC transition at $T_c$(onset) =
21.6 K (90$\%\rho_n$) with a transition width $\Delta T_c$ = 1.7 K
(10\%-90$\% \rho_n$ ), where $\rho_n$ is the resistivity of the
normal state before the SC transition, demonstrating the high
quality of our single crystal samples. The slightly semiconducting
behavior just above $T_c$ along with the $T_c$ value indicates that
the present sample locates on the slightly underdoped region near
the optimal doped point of the phase diagram.

We perform the measurements of temperature dependent electronic
resistivity with the magnetic field along various orientations to
study the anisotropy effect of $H_{c2}$ of superconductor
CaFe$_{1-x}$Co$_{x}$AsF. As shown in Fig.~\ref{fig1}(c) and (d), the
SC transition point shifts to lower temperature with the increase of
the magnetic field for both the orientations of parallel and
perpendicular to $c$-axis of the crystal structure. It is worthy to
note that the SC transition for the orientation of $H//c$ shifts
much quicker than that of $H\bot c$ by comparing the data from
Fig.~\ref{fig1}(c) and (d). Quantitatively, we use $90\%\rho_n$ to
determine the upper critical field $H_{c2}$. The temperature
dependence of $H_{c2}$ is shown in the inset of Fig.~\ref{fig1}(d)
for both the two orientations, and the value of anisotropy parameter
is estimated to be $\gamma = H^{ab}_{c2}/H^{c}_{c2} = 9$ at
temperature of 21.2 K.

\begin{figure}
\includegraphics[width=9cm]{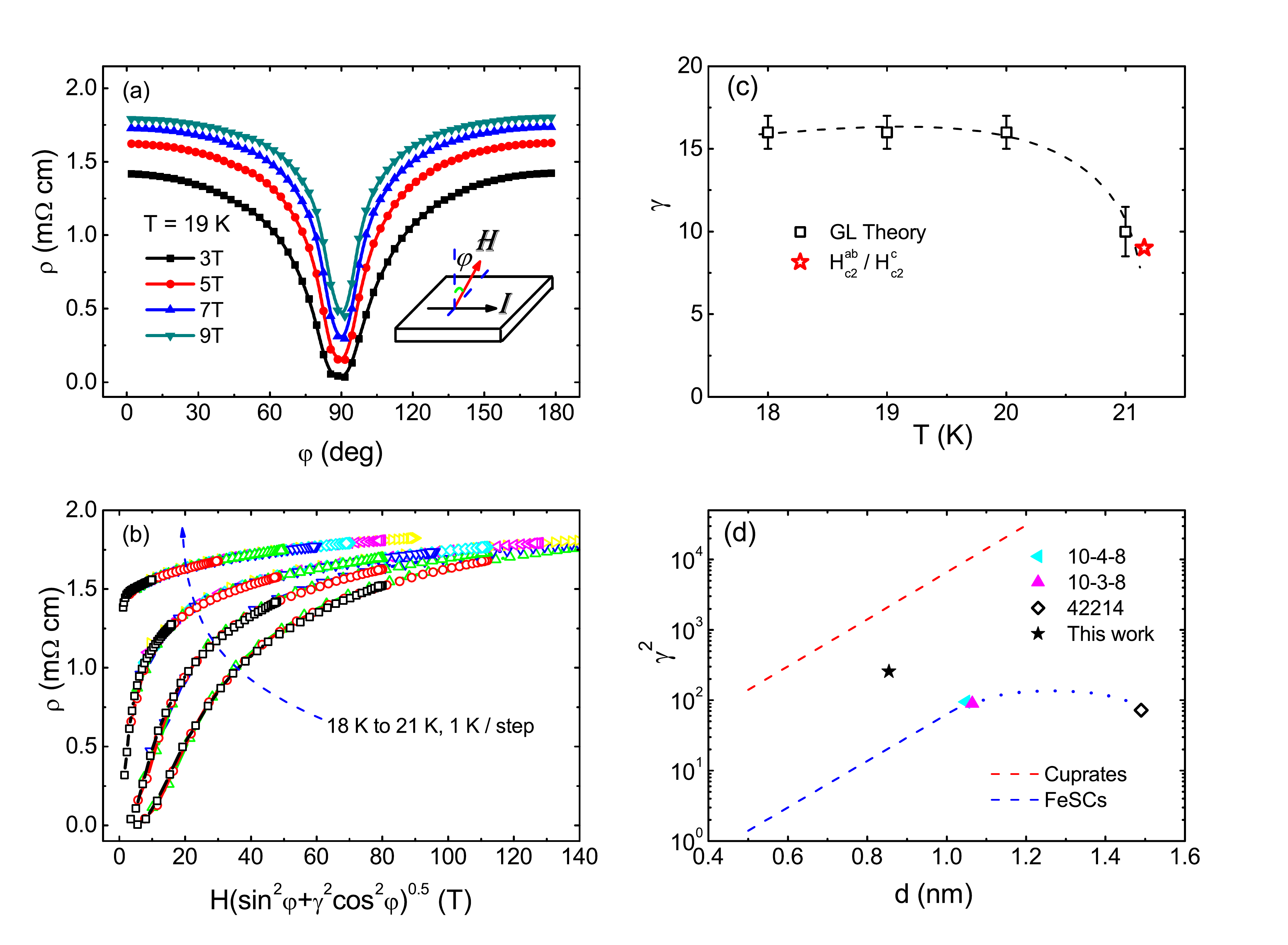}
\caption {(a) Magnetic field angular-dependence of electronic
resistivity at $19$ K under magnetic fields up to 9 T. The inset
illustrates schematically the definition of angle $\varphi$, and the
electron current is always perpendicular to the magnetic field. (b)
Scaled resistivity at 18 K, 19 K, 20 K and 21 K as a function of
$H(\sin^2\varphi + \gamma^2\cos^2\varphi)^{0.5}$. (c) Temperature
dependence of the anisotropy parameters $\gamma$ obtained from the
GL theory and $H^{ab}_{c2}/H^{c}_{c2}$. (d) Comparing our result of
anisotropy with other FeSCs and cuprates. Here $\gamma^2$ is
extracted at the temperature 0.85$T_c$ and $d$ is the distance of
the neighboring conducting layers. The two dashed lines representing
the tendency of the previously reported FeSCs and cuprates
superconductors are cited from  Ref.~[\onlinecite{10-4-8}]. The
value of the 42214 system was reported in Ref.~[\onlinecite{42214}],
which implies a saturated tendency as $d$ becomes even larger.}
\label{fig2}
\end{figure}

To gain more information about the anisotropy effect of $H_{c2}$, we
further measure the magnetic-field-angle dependence of electronic
resistivity under various magnetic fields at the temperature region
near the SC transition point. Here we take the data at the
temperature of 19 K as an example and plot it in Fig.~\ref{fig2}(a).
In addition, the schematic illustration of applied magnetic field
and the dc current is also shown in the inset of Fig.~\ref{fig2}(a).
The positive magnetoresistivity displays a V-shaped structure with
occurring the minimum at $\varphi$ = 90$^{\circ}$ and maximum at
$\varphi$ = 0$^{\circ}$ and $\varphi $ = 180$^{\circ}$. Similar
results are also obtained for other temperatures indicating the
universal behaviours of magnetic-field-angle dependent
magnetoresistivity. Based on the anisotropic GL theory, the
effective upper critical field as a function of the orientation of
azimuth angle $\varphi$ is expressed as~\cite{Blatter1}:
\begin{equation}\begin{split}
H_{c2}^{GL}(\varphi)& = \frac{H_{c2}^{ab}}{\sqrt {{{\sin }^2}\varphi
+ {\gamma ^2}{{\cos }^2}\varphi}} \propto \frac{1}{\sqrt {{{\sin
}^2}\varphi + {\gamma ^2}{{\cos }^2}\varphi}}.\label{eq:1}
\end{split}\end{equation}
Here, it should be noted that the resistivity in the superconducting
state merely depends on the effective magnetic field
$H/H^{GL}_{c2}$, i.e., $\rho=\rho[H/H^{GL}_{c2}(\varphi)]$, at a
given temperature.~\cite{Blatter2} According to the
Eq.~(\ref{eq:1}), we rewrite the resistivity as a function of
$H\sqrt {{{\sin }^2}\varphi + {\gamma ^2}{{\cos}^2}\varphi}$, and
show it in Fig.~\ref{fig2}(b). It clearly demonstrates the universal
behaviors that all magnetic field dependent electronic resistivity
falls into one curve at a fixed temperature. In addition,
temperature dependence of the fitted anisotropic parameter $\gamma$
is also shown in Fig.~\ref{fig1}(c). For a comparison, we also show
a consistent value of anisotropic parameter $\gamma$ estimated from
$H^{ab}_{c2}/H^{c}_{c2}$. Importantly, these anisotropic values are
much larger than that reported on other FeSCs.

In Fig.~\ref{fig2}(d), we also show the $\gamma^2$ values at 0.85
$T_c$ as a function of the distance $d$ between the neighboring
conducting layers (FeAs layers or CuO$_2$ layers). The two dashed
lines, respectively, represent the tendency of the previously
reported FeSCs and cuprates superconductors.~\cite{10-4-8} Generally
speaking, the value of $\gamma^2$ for FeSCs is two orders of
magnitude lower than that of cuprates superconductors with the same
value of $d$. Exceptionally, the value of
CaFe$_{0.882}$Co$_{0.118}$AsF locates above the blue dashed line
which indicates a common tendency of FeSCs, and even higher than the
10-3-8, 10-4-8 and 42214 systems
[Ca$_{10}$(Pt$_n$As$_8$)(Fe$_{2-x}$Pt$_x$As$_2$)$_5$ (n = 3, 4),
Pr$_4$Fe$_2$As$_2$Te$_{1-x}$O$_4$] with a much larger space between
the FeAs layers.~\cite{10-4-8,42214}

\section{Discussion}

Although we have found the largest anisotropy parameter $\gamma$ of
CaFe$_{0.882}$Co$_{0.118}$AsF in iron-based superconductors, the
experimental observation is consistent with the theoretical
prediction by the first-principal calculations. Previously, D. J.
Singh {\it et al}~\cite{Singh1} predicted the anisotropy to be about
15 for the parent phase of the 1111 system after the discovery of
the high-$T_c$ FeSCs. Since the anisotropy is tightly related to the
electronic structure, and as we known that the Fermi surface of
FeSCs consists of five sheets, where two electron-like cylinders
centered around the M-A line and two hole-like cylinders around the
$\Gamma$-Z line, as well as an additional small 3D hole-like pocket
centered at Z point, individual characters in details of the
electronic structure may have remarkable influences on the
anisotropic effect of the materials, such as the strong warping of
the electron pockets in the 122, 111, and 11 systems, compared with
the 1111 system,~\cite{Singh1,Singh2,FeSeTe} can be expected to
increase the tendency of 3D characterizations, which may explain the
fact that the anisotropy parameter of these three systems is smaller
than of the 1111 system.

\begin{figure}
\includegraphics[width=9cm]{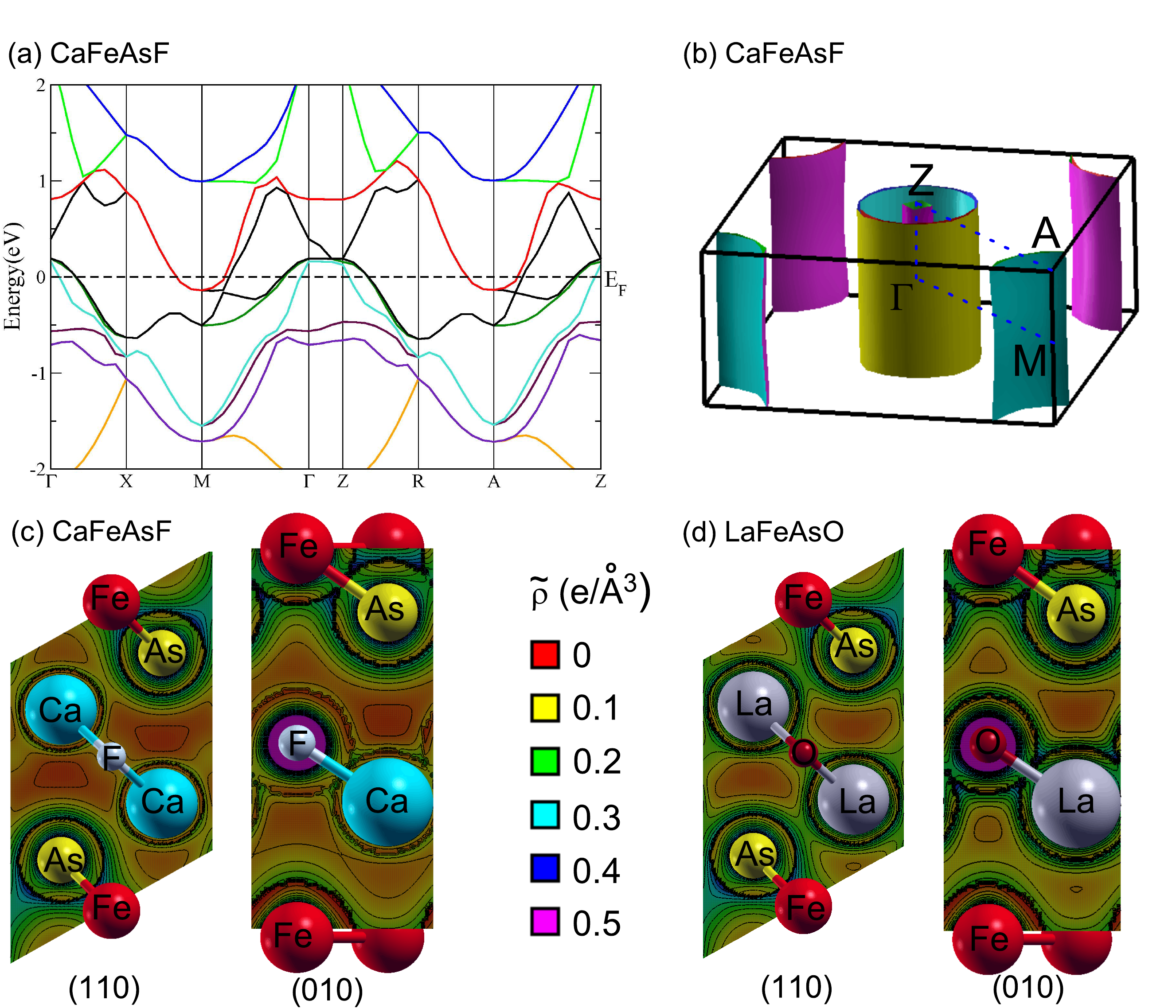}
\caption {(a) Calculated band structure of CaFeAsF. The Fermi energy
was set to zero (dashed line). There is no band intersecting the
Fermi level between the $\Gamma$ and Z points. (b) The Fermi
surfaces of CaFeAsF. (c)-(d) Charge density distribution
($\tilde{\rho}$) of the (110) and (010) planes for CaFeAsF and
LaFeAsO respectively.} \label{fig3}
\end{figure}

In order to examine the detailed electronic structure of
CaFe$_{0.882}$Co$_{0.118}$AsF and clarify the nature of the
emergence of strongest electronic anisotropy among the FeSCs, we
carried out the first-principles calculations. The electronic band
structure should be changed slightly by the cobalt doping of 11.8\%
based on the rigid-band approximation. For simplicity, we just only
show the energy band and its Fermi surface topology of CaFeAsF in
Fig.~\ref{fig3}(a) and (b), respectively, which are in good
agreement with previous report.~\cite{Shein} Firstly we examine the
energy dispersions between high symmetric $\mathbf{k}$-points of
$\Gamma$ and $Z$, there is no band intersects across the Fermi
level. This result indicates the absence of the 3D Fermi pocket
around the $Z$ point. In addition, it is found that all of the Fermi
surface sheets are almost ideal cylinders [see Fig.~\ref{fig3}(b)].
Previously, I. A. Nekrasov {\it et al} have pointed out that the
fluorine-based 1111 system displays a much more perfect 2D character
of the Fermi surfaces than the oxygen-based 1111 system
LaFeAsO~\cite{Nekrasov}. As for the compounds with an even larger
interlayer distance like the 10-3-8 system, both the ARPES
measurements and first-principles calculations have demonstrated a
strong 3D character in the topology of the Fermi surface stemming
from the strong hybridization with the Pt $d_{z^2}$
orbitals.~\cite{10-3-8} This should definitely suppress the
bidimensionality of the Fermi surface. The perfect bidimensionality
of CaFeAsF may explain the experimental fact of the strong
anisotropy of CaFe$_{0.882}$Co$_{0.118}$AsF qualitatively.

Quantatively, we calculated the charge density distribution,
$\tilde{\rho}(r)$, as shown in Fig.~\ref{fig3}(c). For comparison,
we also performed the charge density distribution $\tilde{\rho}(r)$
calculation for the system of LaFeAsO with the same crystal
structure of CaFeAsF and shown in Fig.~\ref{fig3}(d). From the
distribution of $\tilde{\rho}(r)$ in the (110) plane, it displays no
difference of the $\tilde{\rho}(r)$ around Fe-As bond between
LaFeAsO and CaFeAsF, but the $\tilde{\rho}(r)$ in the area of La-As
has clearly much higher value than that in the area of Ca-As,
suggesting the weakness of the interlayer coupling between the
layers of FeAs and CaF. This result is also understandable because
the strong electronegativity of F favors to form the much stronger
ionic binding with Ca, as revealed by the lower values of
$\tilde{\rho}(r)$ in between Ca$^{2+}$ and F$^{-}$ ions in the (010)
plane, and weakens the interlayer coupling. This mechanism well
explains the nature of such a strong electronic anisotropy observed
in experiments in the system of CaFeAsF.

\section{Conclusions}
In summary, magnetic field angular dependence of resistivity in the
superconducting state was measured on the single crystals of
CaFe$_{0.882}$Co$_{0.118}$AsF. The obtained anisotropy parameter
$\gamma$ based on the anisotropic Ginzburg-Landau theory is much
larger than that of other FeSCs. By the charge density distribution
calculations, we found that the strong electronic anisotropy mainly
comes from the strong intra-layer ionic bonding in CaF layer, which
weakens the interlayer coupling between the layers of FeAs and CaF.
Our results demonstrate that the fluorine-base 1111 system may be
the most 2D superconducting material in FeSCs, from the viewpoint of
electronic structure. The interplay between strong electronic
anisotropy and unconventional superconductivity is an important
issue for the understanding the mechanism of unconventional
superconductor, which deserves our attentions in the future.

\begin{acknowledgments}
This work is supported by the National Natural Science Foundation of
China (Nos. 11204338 and 11404359), the ``Strategic Priority
Research Program (B)" of the Chinese Academy of Sciences (Nos.
XDB04040300, and XDB04030000) and Youth Innovation Promotion
Association of the Chinese Academy of Sciences (Nos. 2015187 and
2016215).
\end{acknowledgments}

\end{document}